# Enhanced Hybrid Deep Learning Approach for Botnet Attacks Detection in IoT Environment

A. Karthick kumar[1], S. Rathnamala[2], T. Vijayashanthi[3], M. Prabhananthakumar[4],
Alavikunhu Panthakkan[5], Shadi Atalla[6], and Wathiq Mansoor[7]
[1,2,3]Sethu Institute of Technology, Kariapatti, Virudhunagar, India.
[4]Vellore Institute of Technology, Vellore,Tamil Nadu, India.
[5,6,7]College of Engineering and IT, University of Dubai, U.A.E.
Corresponding Author: karthickdhana@sethu.ac.in; apanthakkan@ud.ac.ae

*Abstract*—Cyberattacks in an Internet of Things (IoT) environment can have significant impacts because of the interconnected nature of devices and systems. An attacker uses a network of compromised IoT devices in a botnet attack to carry out various harmful activities. Detecting botnet attacks poses several challenges because of the intricate and evolving nature of these threats. Botnet attacks erode trust in IoT devices and systems, undermining confidence in their security, reliability, and integrity. Deep learning techniques have significantly enhanced the detection of botnet attacks due to their ability to analyze and learn from complex patterns in data. This research proposed the stacking of Deep convolutional neural networks, Bi-Directional Long Short-Term Memory (Bi-LSTM), Bi-Directional Gated Recurrent Unit (Bi-GRU), and Recurrent Neural Networks (RNN) for botnet attacks detection. The UNSW-NB15 dataset is utilized for botnet attacks detection. According to experimental results, the proposed model accurately provides for the intricate patterns and features of botnet attacks, with a testing accuracy of 99.76%. The proposed model also identifies botnets with a high ROC-AUC curve value of 99.18%. A performance comparison of the proposed method with existing state-of-the-art models confirms its higher performance. The outcomes of this research could strengthen cyber security procedures and safeguard against new attacks.

## I. INTRODUCTION

The development of internet technology revolutionized communication and access to information, leading to its rapid and widespread adoption by the masses. This transformation affected various aspects of daily life. The growing use of Internet technology has both benefited society and brought about a number of bad uses that have an impact on people. These negative aspects include cyberattacks, privacy concerns, social isolation, and mental health issues [1, 2]. Botnet attacks in the IoT environment are a growing concern because of the proliferation of connected devices and the often insufficient security measures implemented on these devices as shown in Fig. 1.

These devices, which can include smart home appliances, industrial machinery, and wearable technology, are afflicted with ransomware that transforms them into bots or zombies. In Fig. 2, The attacker uses this botnet to perform coordinated and harmful activities, such as launching distributed denial-of-service (DDoS) attacks, stealing data, sending spam, or spreading further malware. In particular, IoT devices are susceptible to this kind of attack because their security mechanisms are frequently lacking [3, 4, 5].

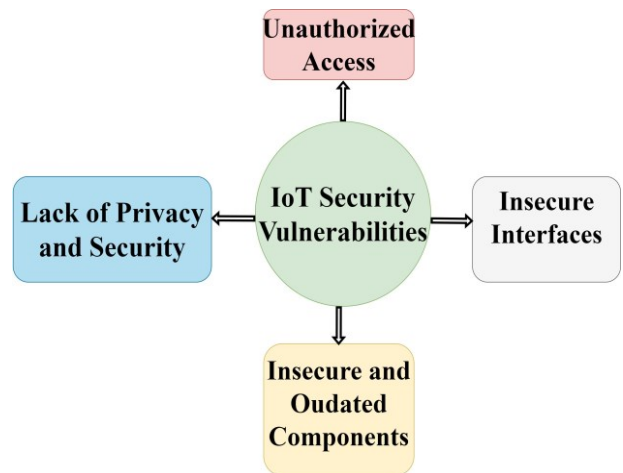

Fig. 1. Security Vulnerabilities in IoT Environment

Botnets can flood IoT devices or services with an overwhelming amount of traffic, rendering them unable to respond to legitimate requests. This can cause critical applications, such as smart grid systems, healthcare monitors, and industrial control systems, to fail, potentially leading to significant operational disruptions [6, 7].

Deep learning-based botnet detection systems are highly effective addressing problems in real-time for enterprises [8]. Hybrid deep learning Approaches can be designed to process massive amounts of data in real time, enabling the timely detection of botnet activities. This is essential for mitigating the impacts of attacks as they occur [8]. This combination improves overall detection accuracy. By implementing deep learning algorithms into botnet detection strategies, researchers can significantly enhance the accuracy and efficacy of detecting botnet activities. These models enable more accurate feature extraction, real-time and predictive analysis, and adaptive learning capabilities, leading to proactive protection strategies [9,10, 11].

The primary contribution of this research as follows:

(i) To strengthen IoT system security, this research proposes a stacking model for botnet attack identification. The hybrid combination of Artificial Neural Networks (ANN), Deep Convolutional Neural Networks (DCNN), Bi-Directional Gated Recurrent Unit (Bi-GRU), Bi-Directional Long Short-Term Memory (Bi-LSTM) are the strengths that are employed in the proposed model.

(ii) Through experiments, many attack types including popular ones like conventional and botnet attacks are classified.

(iii) A comprehensive set of widely recognized performance evaluation metrics, such as recall, accuracy, precision, and the F1 score, ROC-AUC metrics are carefully used to assess the efficacy of the proposed approach.

The preceding research is organized into the following sections: Section II presents a comprehensive literature review, summarizing key research studies and technological advancements relevant to the topic. Section III describes the research methodology, elaborating on the data collection processes and the analytical techniques employed. Section IV introduces the proposed strategy and showcases the findings. Section V analyzes and interprets the results in detail. Finally, Section VI concludes the study, linking the insights gained to the broader field of research and discussing potential future directions.

## II. RELATED STUDIES

Botnets, which are networks of compromised computers controlled by a master host, are employed for malicious applications like spam distribution, DDoS attacks, and data theft. Traditional detection methods, including anomaly-based and signature-based techniques, struggle to identify unknown botnets, handle encrypted traffic, and counter sophisticated evasion tactics used by attackers. Deep learning methods represent a potent avenue for addressing the challenges posed by botnet attacks, offering a sophisticated approach to identifying and mitigating these threats. By leveraging complex neural network architectures and training algorithms, deep learning can effectively analyze large volumes of network traffic data, discerning subtle changes and anomalies indicative of botnet activity.

The CNN-LSTM model was integrated by the authors using the CICIDS 2017 dataset to detect Distributed Denial of Service (DDoS) attacks, as described in reference [12]. Their findings revealed notable performance metrics, with an accuracy rate of 97.16%, and precision reaching 97.41%. The research employs [13] CNN and LSTM networks are leveraged to detect botnet attacks. Specifically, the research evaluates the performance of two models, DNNBoT1 and DNNBoT2, in this task. The comparison of these models reveals that, on average, the training accuracy achieved was notably high. Specifically, the average training accuracy for DNNBoT1 was recorded at 90.71%, while DNNBoT2 exhibited a slightly higher average training accuracy of 91.44%.

In this research [20], an Artificial Neural Network (ANN) model was developed and evaluated using the CTU-13 dataset. When compared to traditional machine learning methods like SVM and NB, the ANN model achieved accuracy of 99%. This indicates that the ANN model is highly efficient for the given dataset, outperforming the SVM and NB models in accuracy. The authors employed domain generation algorithms (DGAs) based on LSTMs to detect botnet attacks. Various DGAs were used, both with benign and DGA domain names. The LSTM model's accuracy for binary classification on two distinct test data sets was 98.7%, while its accuracy scores for multi-class classification were 68.3% and 67.0%, respectively [21]. Table I Describes the existing hybrid deep learning approaches and their inferences.

## III. METHODOLOGY

In this research, a hybrid deep learning approach is proposed for botnet detection. The recommended method is based on a hybrid model in which the final prediction is made using the output of an ANN, DCNN, Bi-LSTM, and RNN. In this research, the performance of proposed model is evaluated concerning their effectiveness in analyzing botnet attack detection using the UNSW-NB15 dataset as described in Figure 3.

### A. UNSW-NB 15 Dataset Description

The UNSW-NB 15 dataset was collected by researchers at the University of New South Wales (UNSW). It was specifically designed to analyze and understand network behavior, providing a robust foundation for studying various aspects of network security, including detecting botnet activities.

Researchers and cybersecurity professionals have extensively utilized the UNSW-NB15 dataset to evaluate the performance of intrusion detection systems. This dataset is pivotal for developing and testing algorithms aimed at identifying various types of network attacks [22]. The UNSW-NB 15 dataset comprises a total of 82,332 records, encompassing various attack types and also including a category labeled as 'Normal', representing benign network behavior.

In Table II describes UNSW NB 15 Dataset nine distinct attack types: 'Generic', 'Exploits', 'Fuzzers', 'DoS' (Denial of Service), 'Reconnaissance', 'Analysis', 'Backdoor', 'Shell code', and 'Worms' [23, 24]. To ensure fairness and effectiveness in training models using the UNSW NB15 dataset, it's imperative to balance the number of instances across all attack types. By equalizing the instances for each attack type, totaling 82,332 cases, the dataset becomes representative and unbiased, enabling more accurate model evaluations across all attack categories.

## IV. PROPOSED HYBRID DEEP LEARNING BOTNET ATTACKS DETECTION MODEL

The Hybrid deep learning Approaches combine various types of neural network architectures or integrate deep learning with other machine learning techniques to leverage the strengths of each approach. Using an ANN+DCNN+BiLSTM+RNN hybrid deep learning model for botnet attack detection offers a multitude of benefits due to the complementary strengths of each neural network type. DCNNs are excellent for extracting spatial features from raw network traffic data. In order to detect abnormalities connected to botnet activity, they are able to recognize complex patterns in header data and packet structures. The BiLSTM and RNN components can adapt to evolving botnet behaviors over time, ensuring that the model remains effective even as attack patterns change. Applying a single model may not yield optimal results; however, combining multiple models through stacking can significantly enhance accuracy. This approach is innovative in the context of these strategies. The specified hybrid deep learning model hyper parameters are described in Table III.

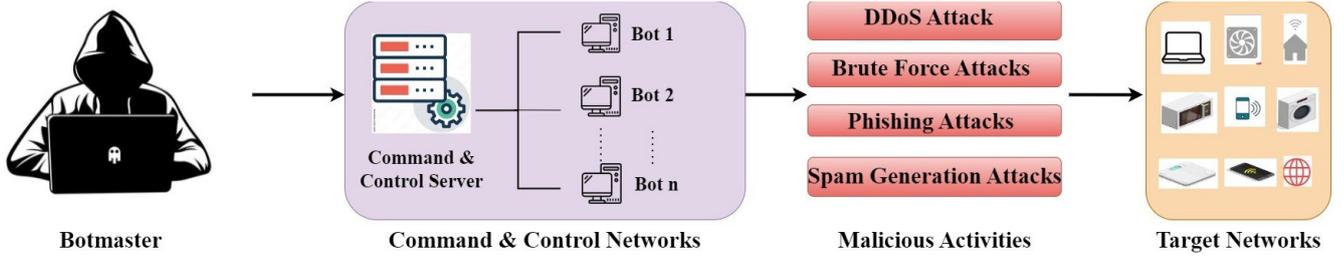

Fig. 2. Botnet Attacks in IoT Environment

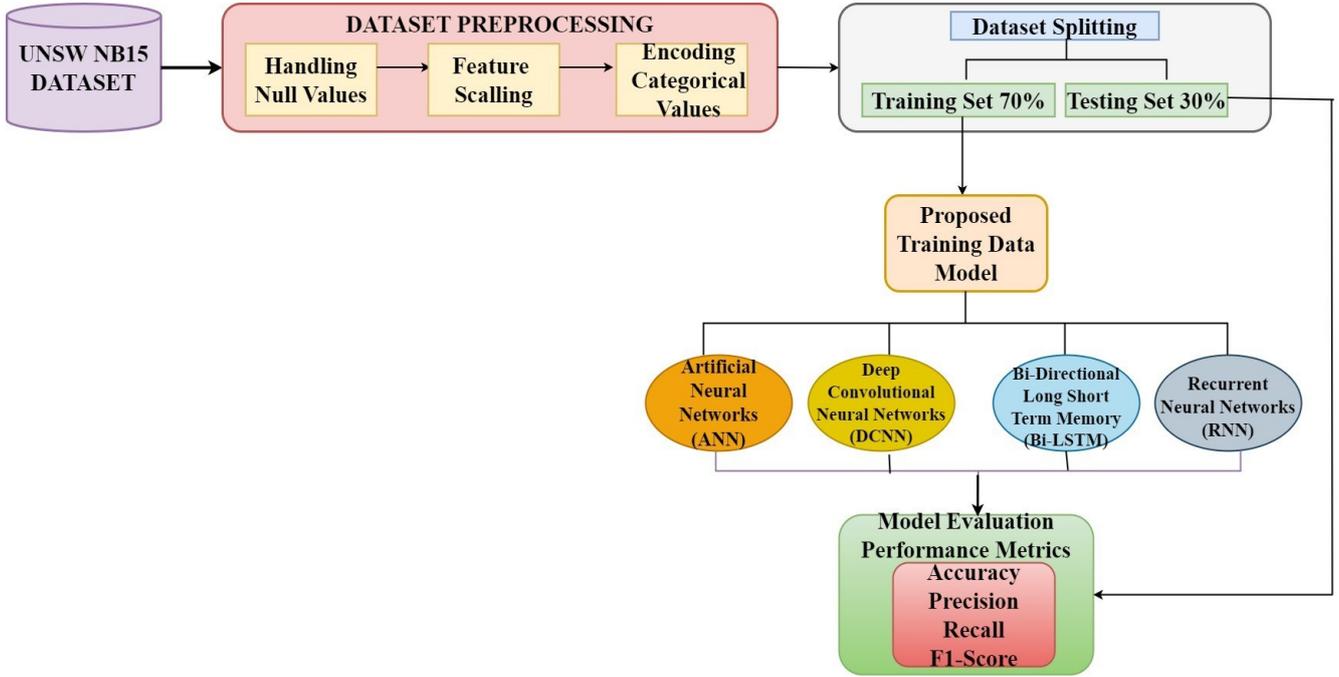

Fig. 3. Proposed Hybrid Botnet Attacks Detection Model

TABLE I: Existing Works

| References | Methodology | Dataset | Inferences | Limitations |
|---|---|---|---|---|
| Rini et al. [14][2024] | CNN-Quantum layers | ALASKA2 | Accuracy 97% | Accuracy improvement |
| Latifah et al [15][2023] | CNN-QRNNN | N-BaIoT | Accuracy 99.43%, Precision 99.13%, Recall 99.12%, | Training time |
| Abdulwahab et al [16][2021] | CNN-LSTM | Smart Grid Dataset | Accuracy 93.00% | Accuracy Improvement |
| Kaiyuan et al [17] [2021] | CNN-BiLSTM | UNSW-NB15 Dataset | Accuracy 85.38% | - |
| Halbouni et al [18][2022] | CNN-LSTM | SUNSW-NB15 Dataset | Accuracy 94.53% | Detection Rate |
| Muhammad et al [19][2022] | PP-DIP | SUNSW-NB15 Dataset | Accuracy 96.30% | -. |

TABLE II: The UNSW-NB 15 Dataset Descriptions

| Sl.no | Attacks types | Descriptions |
|---|---|---|
| 1 | Normal | Natural Data Interactions |
| 2 | Fuzzers | Random data sent to applications to find vulnerabilities |
| 3 | Backdoors | Unauthorized access gained by bypassing normal authentication |
| 4 | Denial of Service | Flooding resources to make them unavailable to users |
| 5 | Exploits | Taking advantage of vulnerabilities for unauthorized access |
| 6 | Generic | Non-specific attacks exploiting general weaknesses |
| 7 | Reconnaissance | Gathering information to identify vulnerabilities |
| 8 | Analysis | Techniques for network traffic analysis and reconnaissance |
| 9 | Shellcode | Injecting and executing malicious code |
| 10 | Worms | Self-replicating malware spreading across networks |

TABLE III PROPOSED MODEL HYPERPARAMETER SETTING

| Models | layers | Unit | Activation Function | Optimizer | Epoch | Accuracy(%) |
|---|---|---|---|---|---|---|
| ANN | 3 | 64,32,1 | relu, tanh | adagrad | 20 | 80.25 |
| ANN | 3 | 64,32,1 | softmax, sigmoid | sgd | 20 | 72.21 |
| ANN | 3 | 64,32,1 | softmax, relu | adam | 20 | 73.89 |
| CNN | 6 | 64,32 | tanh, relu | adagrad | 30 | 93.55 |
| CNN | 6 | 64,32 | softmax, relu | rmsprop | 30 | 80.25 |
| CNN | 6 | 64,32 | tanh, relu | adam | 30 | 91.27 |
| LSTM | 2 | 64,1 | sigmoid | adamax | 30 | 95.03 |
| LSTM | 2 | 64,1 | relu | adamax | 30 | 94.57 |
| LSTM | 2 | 64,1 | tanh | rmsprop | 30 | 96.88 |
| RNN | 2 | 64,1 | softmax | sgd | 25 | 93.38 |
| RNN | 2 | 64,1 | tanh | rmsprop | 25 | 91.25 |
| RNN | 2 | 64,1 | sigmoid | adam | 25 | 92.58 |
| **Proposed model** | **18** | **64,32,1** | **relu, sigmoid** | **adagrad** | **25** | **99.76** |
| Proposed model | 18 | 64,32,1 | tanh, sigmoid | adamax | 25 | 95.41 |
| Proposed model | 18 | 64,32,1 | relu, tanh | adam | 25 | 93.37 |

TABLE V COMPARATIVE ANALYSIS OF ALL MODELS

| References | Existing models | Dataset | Classes | Accuracy (%) | Precision (%) | Recall (%) | F1-Score (%) |
|---|---|---|---|---|---|---|---|
| [25] | LSTM | UNSW-NB 15 | Binary class | 70.00 | - | - | - |
| [26] | CNN-LSTM | UNSW-NB 15 | Binary class | 93.68 | 94.63 | - | - |
| [27] | CNN1D | UNSW-NB 15 | Binary class | 89.80 | - | - | - |
| [28] | CNN-LSTM | UNSW-NB 15 | Binary class | 89.93 | 86.15 | - | 90.43 |
| **Proposed model[ANN-DCNN-BiLSTM-RNN]** | - | **UNSW NB 15** | **Multiclass class** | **99.76** | **98.46** | **97.10** | **98.66**. |

## V. RESULT AND DISCUSSIONS

When evaluating a hybrid deep learning model, it's essential to consider various metrics to assess its performance accurately.

$$\text{Accuracy} = \frac{\text{TrP} + \text{TrN}}{\text{TrP} + \text{TrN} + \text{FrP} + \text{FrN}} \quad (1)$$

$$\text{Precision} = \frac{\text{TrP}}{\text{TrP} + \text{FrP}} \quad (2)$$

$$\text{Recall} = \frac{\text{TrP}}{\text{TrP} + \text{FrN}} \quad (3)$$

$$\text{F1-Score} = \frac{2 \times (\text{Precision} \times \text{Recall})}{\text{Precision} + \text{Recall}} \quad (4)$$

$$\text{ROC-AUC} = \int_0^1 \text{TPR}(\text{FPR}) \, d(\text{FPR}) \quad (5)$$

In this study, we compare the most popular approaches for deep learning categorization. Additionally, we evaluated the suggested ANN+DCNN+Bi-LSTM+RNN model's performance. The experimental findings using the suggested model are shown in Table IV. Table V highlights the comparative results of all models employed in this research, focusing on accuracy, precision, recall, and F1 scores.

The ANN+DCNN+Bi-LSTM+RNN model shows enhanced performance over other models being fine-tuned for optimal outcomes with the given dataset. The ROC and AUC are important metrics used for evaluating the performance of proposed model. Fig. 4 shows the ROC-AUC score for the proposed model achieved 99.18%.

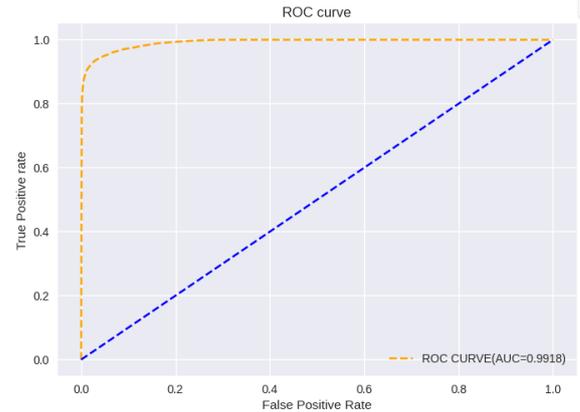

Fig. 4. Performance of ROC-AUC Curve

TABLE IV: PERFORMANCE OF PROPOSED MODEL

| Epoch | Accuracy | Precision | Recall | F1-Score | Training time (s) |
|---|---|---|---|---|---|
| 5 | 96.35 | 94.66 | 97.10 | 95.77 | 967.5 |
| 10 | 96.11 | 95.27 | 96.01 | 95.63 | 1754 |
| 15 | 97.36 | 96.74 | 95.99 | 98.66 | 2502 |
| 20 | 98.42 | 97.79 | 95.89 | 97.16 | 3647.5 |
| **25** | **99.76** | **98.46** | **97.10** | **98.66** | **4643.5** |

## VI. CONCLUSION

The Botnet attacks can facilitate unauthorized access to networks and devices, leading to data breaches and the theft of sensitive information. Deep learning models have significantly enhanced the identification of botnet attacks, providing several advantages over traditional methods. This research proposes an innovative hybrid stacking model, designated as ANN+DCNN+Bi-LSTM+RNN, specifically designed to detect botnet activities. This advanced approach leverages the strengths of multiple deep learning architectures to enhance detection accuracy and resilience against sophisticated cyber threats. The proposed model performs exceptionally well, as evidenced by its 99.76% accuracy, 98.46% precision, 97.10% recall, and 98.66% F1-Score. In future research, there will be a heightened emphasis on leveraging artificial AI techniques for training, driven by their potential to significantly enhance effectiveness in various domains, particularly in the realm of botnet attack detection.